\documentclass[a4paper,fleqn,usenatbib]{mnras}

\usepackage{newtxtext,newtxmath}

\usepackage[T1]{fontenc}
\usepackage{ae,aecompl}

\usepackage{graphicx}	
\usepackage{amsmath}	
\usepackage{amssymb}	
\usepackage{xspace}
\usepackage{subfig}
\usepackage{tabularx}
\usepackage{hyperref}

\newcommand{\sgx}{\textsc{s}g\textsc{xb}\xspace}

\newcommand{\sg}{Sg\xspace}
\newcommand{\co}{CO\xspace}
\newcommand*{\hmxb}{\textsc{hmxb}\@\xspace}

\newcommand*{\eg}{e.g.\@\xspace}
\newcommand*{\ie}{i.e.\@\xspace}
\newcommand*{\aka}{a.k.a. \@\xspace}

\title[Accretion of a clumpy wind in SgXB]{Accretion from a clumpy massive-star wind\\in Supergiant X-ray binaries}

\author[I. El Mellah, J.O. Sundqvist and R. Keppens]{
I. El Mellah,$^{1}$\thanks{E-mail: ileyk.elmellah@kuleuven.be}
J.O. Sundqvist,$^{2}$
R. Keppens$^{1}$
\\
$^{1}$Centre for mathematical Plasma Astrophysics, Department of Mathematics, KU Leuven, Celestijnenlaan 200B, B-3001 Leuven, Belgium\\
$^{2}$KU Leuven, Instituut voor Sterrenkunde, Celestijnenlaan 200D, B-3001 Leuven, Belgium
}

\date{Accepted XXX. Received YYY; in original form ZZZ}

\pubyear{2017}

\begin{document}
\label{firstpage}
\pagerange{\pageref{firstpage}--\pageref{lastpage}}
\maketitle

\begin{abstract}
Supergiant X-ray Binaries (\sgx) host a compact object, often a neutron star (NS), orbiting an evolved O/B star. Mass transfer proceeds through the intense line-driven wind of the stellar donor, a fraction of which is captured by the gravitational field of the NS. The subsequent accretion process onto the NS is responsible for the abundant X-ray emission from \sgx. They also display peak-to-peak variability of the X-ray flux by a factor of a few 10 to 100, along with changes in the hardness ratios possibly due to varying absorption along the line-of-sight. We use recent radiation-hydrodynamic simulations of inhomogeneities (aka clumps) in the non-stationary wind of massive hot stars to evaluate their impact on the time-variable accretion process. For this, we run 3D hydrodynamic simulations of the wind in the vicinity of the accretor to investigate the formation of the bow shock and follow the inhomogeneous flow over several spatial orders of magnitude, down to the NS magnetosphere. In particular, we show that the impact of the wind clumps on the time-variability of the intrinsic mass accretion rate is severely tempered by the crossing of the shock, compared to the purely ballistic Bondi-Hoyle-Lyttleton estimation. We also account for the variable absorption due to clumps passing by the line-of-sight and estimate the final effective variability of the column density and mass accretion rate for different orbital separations. Finally, we compare our results to the most recent analysis of the X-ray flux and the hardness ratio in Vela X-1.
\end{abstract}




\section{Introduction}

Massive, isolated supergiant O/B stars (\sg) have been thoroughly investigated in the last years under several complementary points of view : using asteroseismology \citep{Aerts2017}, investigating rotation properties and chemical abundance patterns \citep[\eg][]{Martins2016}, and not the least by analyses of their powerful line-driven wind outflows \citep[see review by][]{Puls2008}. But while such studies have brought many significant insights regarding the evolution of massive stars, the recent detection of gravitational waves from coalescing compact objects \citep[\co ,][]{Abbott2016,Abbott2016a} have urged even more the community to evaluate the impact of binarity on the evolutionary tracks.

In particular, we expect to find good candidates as progenitors of compact binaries among the High Mass X-ray Binaries (\hmxb), systems where a stellar-mass black hole or a neutron star (NS), accretes a fraction of the wind emitted by the stellar companion. This mass transfer mechanism called wind accretion is much less constrained than the one in Low Mass X-ray Binaries, where it essentially proceeds through Roche Lobe Overflow \citep{Paczynski1971}. Until now, the Bondi-Hoyle-Lyttleton (BHL) sketch \citep[see the review by ][]{Edgar:2004ip} applied to a smooth wind is used to estimate the mass accretion rate onto the compact object \citep{Davidson1973}. However, the underlying idealized representation, both of the wind and of the accreted flow, might jeopardize any attempt to accurately account for wind mass transfer in stellar evolutionary codes.

\begin{figure*}
\centering
\includegraphics[width=2\columnwidth]{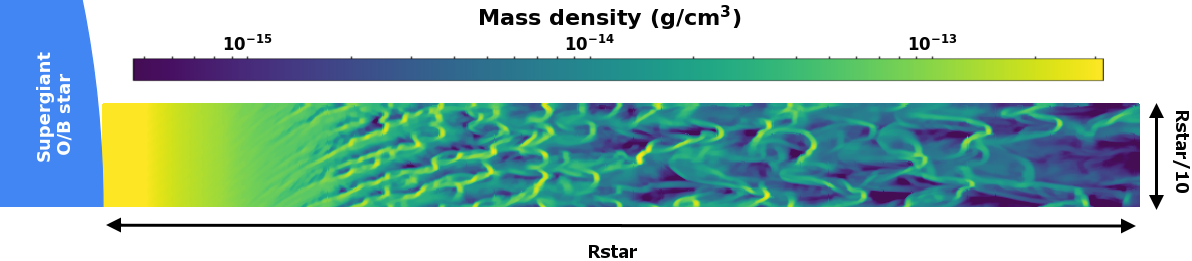}
\caption{Logarithmic color map of the mass density in the wind at a given time, once the flow has reached a statistically steady state \citep{Sundqvist2017}. An animated version can be found at \href{http://homes.esat.kuleuven.be/~ileyk/jon_sundqvist_wind_2D.gif}{http://homes.esat.kuleuven.be/$\sim$ileyk}.}
\label{fig:snapshot}
\end{figure*} 

In the sub-class of classic Supergiant X-ray Binaries (\sgx), the evolved \sg companion undergoes dramatic mass losses through a fast, dense and isotropic line-driven wind : because of resonant line absorption of ultra-violet photons by partly ionized metal ions, the outer stellar layers are provided net outwards linear momentum \citep{Lucy1970,Castor1975}. Such line-driven winds are known to be the siege of an intrinsic and remarkably strong instability \citep[the line-deshadowing instability, LDI,][]{Lucy1980,Owocki1984} whose development leads to internal shocks within the wind. The non-linear growth of this instability has been computed first in one-dimensional simulations \citep[see \eg][]{Owocki1988,Feldmeier1997} but it is only in the 2000's that the first multi-dimensional simulations were performed \citep{Dessart2005}. These first multi-dimensional simulations witnessed a transverse fragmentation of the shells into higher density regions called "clumps", but could not properly resolve the lateral extension of the clumps. Using a two-dimensional pseudo-planar grid sampling a restricted angular region, \cite{Sundqvist2017} (hereafter S17) recently managed to capture the dimensions of the clumps in both directions and to follow them up to two stellar radii. Their results agree with expectations from linear analysis \citep{Rybicki1990} and are consistent with observations \citep{Oskinova2007,Furst2010}. Investigating the way this structured wind can be beamed and captured by the gravitational field of the orbiting NS in a \sgx is the object of the present paper.

Previous numerical models of wind of a massive star with accretion onto an orbiting compact object have been made in \cite{Blondin1990} and \cite{Manousakis2015c}. The authors used a global modeling approach of the wind and the compact object, accounting for the X-ray ionizing feedback on the flow, which fully ionizes the metal ions and subsequently inhibits the acceleration \citep{Hatchett1977}. They also included the orbital effects, twisting and shearing the wind, but considered a smooth wind (suppressing the LDI by using a line-force based on the so-called Sobolev approximation). On the contrary, \cite{Oskinova2012} and \cite{Bozzo2016} modeled the clumps but they relied on the LDI simulations by \cite{Feldmeier1997} which were uni-dimensional and thus did not resolve the transverse structure of the clumps ; the authors warned that the simple BHL analytic recipe for accretion then yields excessive variability, unconsistent with the observations of \sgx. More recently, \cite{Cruz-Osorio2017} worked out the influence of rigid and randomly distributed static spheres on the accretion of a planar supersonic homogeneous BHL flow in a two-dimensional spherical slice where invariance around the polar axis is assumed. \cite{Ducci2009} adopted a statistic representation of spherical clumps in the wind to evaluate their impact on the time-variability of the mass accretion rate. However, they did not solve the dynamical problem and resorted on the BHL prescription to deduce the mass accretion rate as clumps intersect the transverse section of the accretion cylinder. 

In the present paper, we consider a physically-motivated three-dimensional wind derived from full radiation-hydrodynamic simulations. Its dynamics is not altered by the presence of a neutron star (NS) companion up to a few accretion radii around the accretor. Then, we solve the equations of hydrodynamics to compute the evolution of the inhomogeneous flow as it is beamed by the gravitational field of the NS within a few accretion radii and down to a few hundred times the size of the NS (\ie at the outer edge of its magnetosphere). For different orbital separations and mass of the accretor, we compute, as a function of time, the mass accretion rate $\dot{M}$, and the column density $N_\mathrm{H}$ in any direction. We use those quantities to estimate the impact of a realistic wind clumpiness on the time variability of the absorption and X-ray luminosity.

In Section\,\ref{sec:3D_Jon}, we describe how we used the most recent simulations on the non-linear growth of clumps. We then use this wind as outer upstream boundary condition for a numerical setup centered on the accretor and described in more detail in Section\,\ref{sec:num_set} and the appendices. The obtained results on $\dot{M}$ and $N_\mathrm{H}$ are discussed in Section\,\ref{sec:res} and conclusions are summarized in Section\,\ref{sec:conc}.

\section{Clumpy wind representation}
\label{sec:3D_Jon}

Numerical radiation-hydrodynamic simulations of a line-driven wind generally show that the non-linear growth of the LDI leads to high-speed rarefactions that steepen into strong reverse shocks, whereby the wind plasma becomes compressed into spatially narrow "clumps" separated by large regions of rarefied gas \citep{Owocki1988,Sundqvist2013,Feldmeier2017}. In previous 1D models, such clumps really are radial shells ; S17 recently performed 2D radiation-hydrodynamic simulations within 2 stellar radii ($2R_{*}$) which demonstrate how these characteristic LDI-shells are broken up by the influence of oblique radiation rays and basic hydrodynamic instabilities. The width of the two dimensional simulation box, $R_{*}/10$, is large enough to capture the transverse extension of the clumps, while still small enough to work in a pseudo-planar approach. After a few wind dynamical flow time-scales, the multi-dimensional structures develop into a complex but statistically quite steady flow, characterized by localized high-density clumps of small spatial extent embedded in larger regions of much lower density. The full numerical setup is described in detail in S17.

%

These simulations displayed a transition from an initially smooth analytic profile to a statistically steady state. As described above, the transition is due to the LDI \citep{Owocki1984} whose growth first forms shells due to the internal shocks before lateral fragmentation occurs under the action of the lateral component of the line-force. The obtained overdensities range from being sometimes roughly bow-shaped to sometimes more isotropic, have typical sizes of a few $R_{*}/100$ at $2R_{*}$ and are denser than the inter-clump environment by a factor up to a few 100 (see Figure\,\ref{fig:snapshot}). Their total mass is $\sim10^{17-18}$g, available to enhance the permanent mass accretion rate if such a clump would happen to pass close enough to the accretor. Those dimensions and masses are consistent with the observations \citep{Furst2010,Leutenegger2013,Grinberg2015} and theoretical estimates \citep{Ducci2009}, and they favor the lower end of the previously claimed values.

S17 considered an O star with parameters given in Table\,\ref{tab:param}. The main difference between the star they considered and the typical Supergiant O/B host star in a classical \sgx (such as HD 77581 in Vela X-1) is mostly the effective liberation speed at the photosphere, 20\% lower in the current case. It has been shown that the predicted structure of the wind should be little altered by reasonable changes in those parameters.


In this paper, we use the results from S17 as outer boundary conditions to our simulation space centered on the accretor. We extend the results to 3D while still retaining the stochastic properties of the flow (histograms and correlation map) following the methods presented in more detail in Appendix\,\ref{app:3Drec}.

\section{Numerical model}
\label{sec:num_set}

The numerical mesh we work with is a three dimensional extension of the one introduced in \cite{ElMellah2015}, a spherical stretched AMR mesh centered on the accretor whose outer spatial extension is described in Section\,\ref{sec:Racc} below. Within this space, we solve the conservative equations of hydrodynamics using the \texttt{MPI-AMRVAC} finite volume code \citep{Porth:2014wv}. To inject the highly supersonic clumpy wind described in the previous section in the simulation space, we incorporate time-variable outer boundary conditions (Section\,\ref{sec:outBC}). For a more thorough description of the numerical setup itself, we refer to Appendix\,\ref{app:num_divers}.

\subsection{Self defined accretion radius}
\label{sec:Racc}

To avoid any spurious reflection at the outer boundary downstream, once the flow has passed the accretor, we must make sure that the flow is supersonic once it reaches the outer edge \citep{Blondin2009}. The incoming wind is largely supersonic but gravitational beaming of streamlines with an impact parameter smaller than the accretion radius, leads to the formation of a detached bow shock \citep{ElMellah2015} with $R_\mathrm{acc}$ given by :
\begin{equation}
\label{eq:racc}
R_\mathrm{acc}=\frac{2GM}{v_{\infty}^2}
\end{equation}
where $G$ is the gravitational constant, $M$ the mass of the accretor and $v_{\infty}$ the velocity of the flow before it is significantly altered by the gravitational field of the accretor\footnote{Not to be confused here with the wind terminal speed $v(r=\rightarrow \infty)$.} \citep[see eg][]{Edgar:2004ip}. In the wake of the accretor (\aka accretion trail), the flow which manages to escape accretion remains subsonic for a few accretion radii before being supersonic again \citep{Blondin2009}. A spherical simulation space with an outer radius of $8R_\mathrm{acc}$ is enough to witness the formation of the shock without numerical artifacts due to the outer boundary conditions downstream which can simply be taken as continuous.

We compute $R_\mathrm{acc}$ in a low and high orbital separation configuration (respectively the close and wide configurations in Table\,\ref{tab:param}). The effective velocity at the orbital separation, $v_{\infty}$, is computed using the values of radial speed from S17 : once averaged transversally and in time around the orbital separation, $a$, we obtain a speed representative of the amount of specific kinetic energy provided to the wind by the line-force process before it starts to experience the gravitational grasp of the accretor. It serves as the normalization speed. We validated a posteriori that this estimate is valid and does give a suitable outer radial extension to the simulation space. We tune the NS mass so as to work with a similar physical accretion radius in both configurations. For realistic values of the mass of the accretor, of the wind speed at the orbital separation and of the stellar radius, the outer radius of the simulation space is a few times larger than the width of the simulation stripes in our LDI wind models, hence the need for the transverse prolongation described in Appendix\,\ref{sec:trans_prol}. Furthermore, this sphere of radius $8R_\mathrm{acc}$ centered on the accretor coincides approximately with the Roche lobe of the compact object, which justifies to discard the gravitational influence of the compact object outside this region and to rely on the wind simulation from S17.


\begin{table}
\centering
\caption{Parameters of the two configurations considered}
\label{tab:param}
\begin{tabularx}{\linewidth}{c|c|c}
  Parameters & Close config. & Wide config. \\
  \hline
  Stellar mass & \multicolumn{2}{c}{50M$_{\odot}$} \\
  \hline
  Stellar radius & \multicolumn{2}{c}{20R$_{\odot}$} \\
  \hline
  Mass loss rate & \multicolumn{2}{c}{1.3$\cdot$10$^{-6}$M$_{\odot}\cdot$yr$^{-1}$} \\
  \hline
  Orbital separation & 1.6R$_{*}$  & 2R$_{*}$  \\
  \hline
  NS mass & 1.3M$_{\odot}$ & 2M$_{\odot}$  \\
  \hline
  Effective velocity at NS & 925 km$\cdot$s$^{-1}$ & 1140 km$\cdot$s$^{-1}$  \\
  \hline
  Density at orb. sep. & 7.2$\cdot$10$^{-15}$g$\cdot$cm$^{-3}$ & 3.6$\cdot$10$^{-15}$g$\cdot$cm$^{-3}$ \\
  \hline
  Accretion radius & \multicolumn{2}{c}{$\sim 4\cdot 10^{10}$cm $\sim 0.6$R$_{\odot}$} \\
\end{tabularx}
\end{table}


\subsection{Ionization surface}
\label{sec:ion}
\begin{figure}
\centering
\includegraphics[width=0.9\columnwidth]{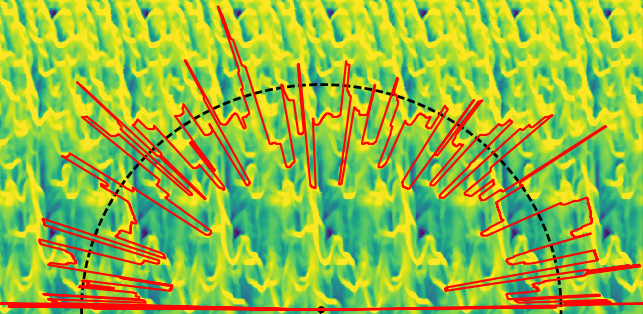}
\caption{Ionization surface in red for a X-ray source luminosity of $2\cdot 10^{36}$erg$\cdot$s$^{-1}$ and a critical ionization parameter of 500. The wind comes from the top, the X-ray source lies at the center of the lower end of the plot and the colormap represents the mass density. The dashed black line represents the extension of the simulation space ($8R_\mathrm{acc}$, see Section\,\ref{sec:Racc}).}
\label{fig:ion}
\end{figure} 

What about the radiative influence of the X-rays emitted in the immediate vicinity of the accretor on the inflowing wind? The wind launching mechanism derived by \cite{Lucy1970} and \cite{Castor1975} for hot stars relies fully on the capacity of partly ionized metal ions to resonantly absorb ultra-violet photons. If those ions get to be fully ionized, no radiative momentum can be tapped by absorption and the acceleration of the wind is suddenly inhibited, an effect first reported by \cite{Hatchett1977}. The ionization parameter defines locally the state of ionization of the flow for an optically thin gas in local ionization and thermal balance :
\begin{equation}
\zeta = \frac{L_\mathrm{X}}{nr^2}
\end{equation}
where $L_\mathrm{X}$ is the X-ray luminosity emitted by the accreting source, $r$ the distance to this source and $n$ the local atomic number density. The exact value of the critical ionization parameter $\zeta_{cr}$ above which the line-force is significantly quenched has been a matter of debate, varying from $10^4$ erg$\cdot$cm$\cdot$s$^{-1}$ \citep{Hatchett1977,Ho1987} to $10^{2.5}$ erg$\cdot$cm$\cdot$s$^{-1}$ \citep{Blondin1990,Ducci2010a,Manousakis2015c} and even lower values \citep{Stevens1991,Krticka2015}. \cite{Oskinova2012} also showed that accounting for clumpiness lowers the ionization parameter compared to a smooth wind, by a factor of $\sim2-3$ for clumping factors (aka wind inhomogeneity parameter) such as those obtained in our LDI simulations.

To evaluate the extension of this region in the upstream hemisphere of the region centered on the compact object, we compute the ionization parameter along a set of rays emitted from the accretor. The computation stops when $\zeta$ reaches $\zeta_{cr}$, which enables us to locate the ionization front for different ratios $L_\mathrm{X}/\zeta_{cr}$ (Figure\,\ref{fig:ion}). Assuming optically thin conditions \citep{Oskinova2012,Grinberg2015,Martinez-Nunez2017}, $\zeta$ describes entirely the local ionization state of the flow. Figure\,\ref{fig:ion} is plotted for $L_\mathrm{X}=2\cdot 10^{36}$erg$\cdot$s$^{-1}$, consistent with the low-luminosity solution of \cite{Ho1987}, and $\zeta_{cr}=500$. Those values are representative of \sgx. Depending on the directions and the time snapshot considered, the ionization front (red line in Figure\,\ref{fig:ion}) approximately matches the outer rim of the simulation space (black dashed arc in Figure\,\ref{fig:ion}). As a first approximation, we assume that they exactly coincide at any time, which means that the wind acceleration is fully inhibited within the simulation space and that the flow is unaltered downstream this region. Consequently, we can inject the clumps at the outer rim of the upstream hemisphere of the simulation space by using the time-varying values of mass density, speed and pressure from the LDI clumpy wind results of S17. In the present paper, we do not account for the time-varying X-ray ionizing feedback from the immediate vicinity of the NS on the extension of this ionized region.


\subsection{Outer boundary conditions}
\label{sec:outBC}

Once we set the orbital separation and the mass of the accretor, we can place the compact object within the wind and estimate the radial extension of the simulation space. Based on the three dimensional reconstruction described in Appendix\,\ref{app:3Drec}, we can determine the time varying outer boundary conditions downstream for the mass density, the velocity and the pressure. To do so, we sample the spherical mesh embedded in the Cartesian three dimensional data and assign values of the closest pixel. Such a straightforward interpolation method guarantees that we do not lower the contrasts determined by the LDI wind simulations. To be sure to resolve the clumps, in particular those with a small impact parameter, likely to be accreted, we realize this sampling with different mesh resolutions corresponding to the different Adaptive Mesh Refinement (AMR) levels described in Appendix\,\ref{app:num_divers}. An animation of the clumps entering the simulation space from the upstream hemisphere can be found at \href{http://homes.esat.kuleuven.be/~ileyk/clumps_HR.gif}{http://homes.esat.kuleuven.be/$\sim$ileyk}.

\begin{figure}
\centering
\hspace*{-0.5cm}
\includegraphics[width=1.1\columnwidth]{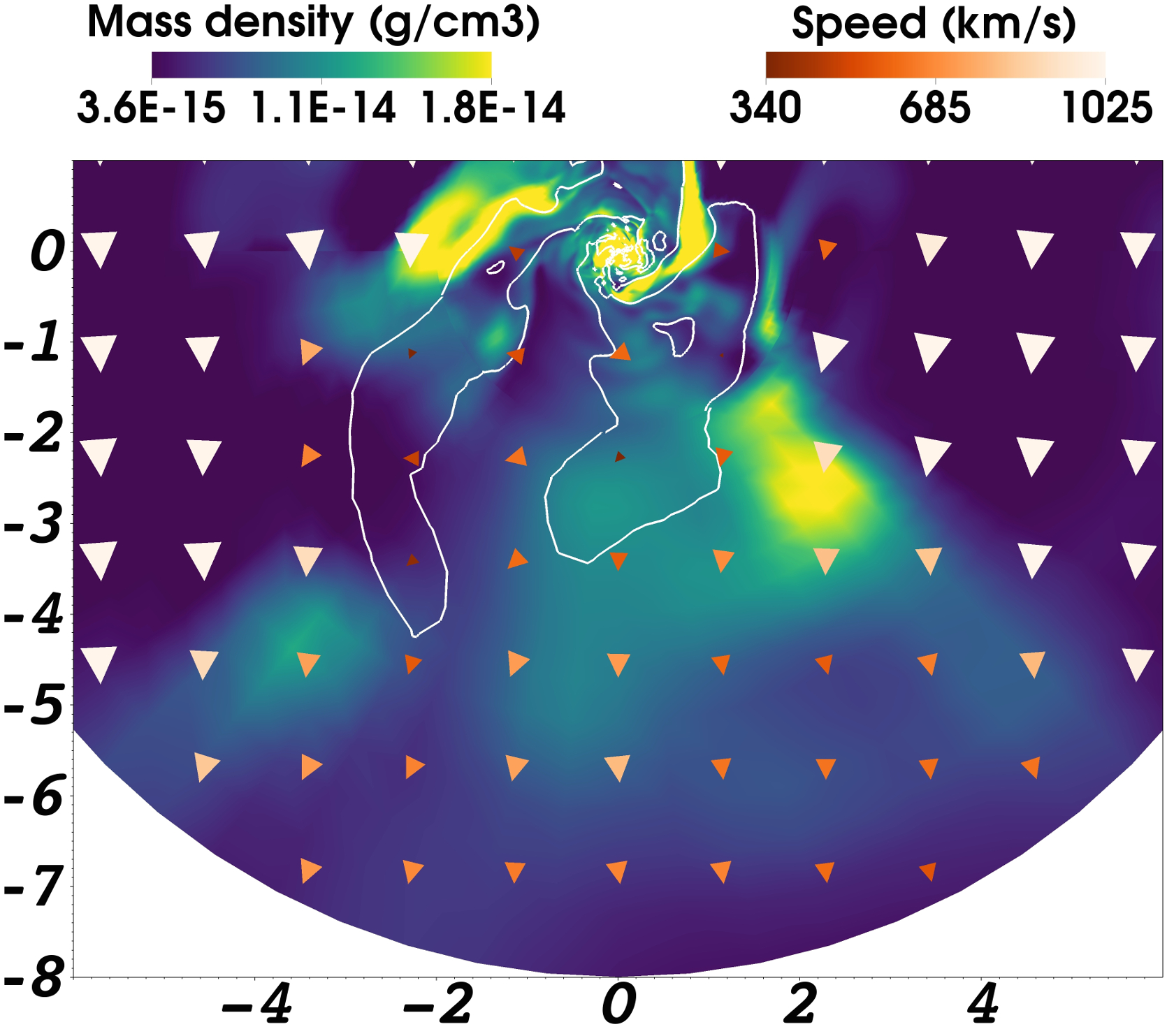}
\caption{Colormap of the density in the equatorial plane of the spherical mesh. We centered the view on the wake and saturated the density color scale to focus on the variations within the wake, with a factor of at least 5 between the dark blue and the yellow regions. The direction and intensity of the velocity field has also been plotted to show the drop in speed (represented by the color and the size of the arrows) and the deviation at the shock. The Mach-1 contour is the white solid line and the length axis are in units of $R_\mathrm{acc}$ (with the accretor at the origin).}
\label{fig:tail}
\end{figure}

\begin{figure*}
  \captionsetup[subfigure]{labelformat=empty}
     \subfloat[]{%
       \includegraphics[width=0.49\textwidth]{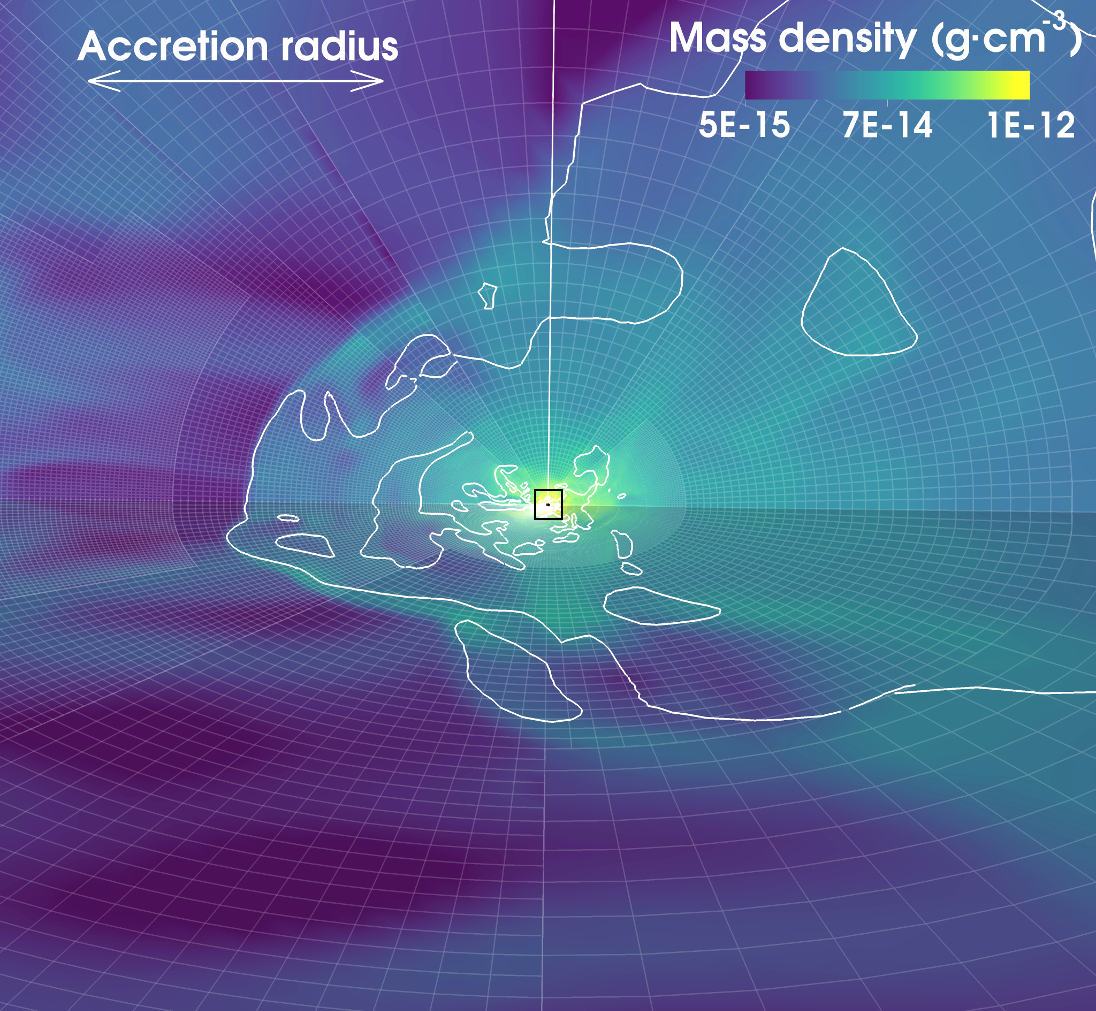}
     }
     \hfill
     \subfloat[]{%
       \includegraphics[width=0.49\textwidth]{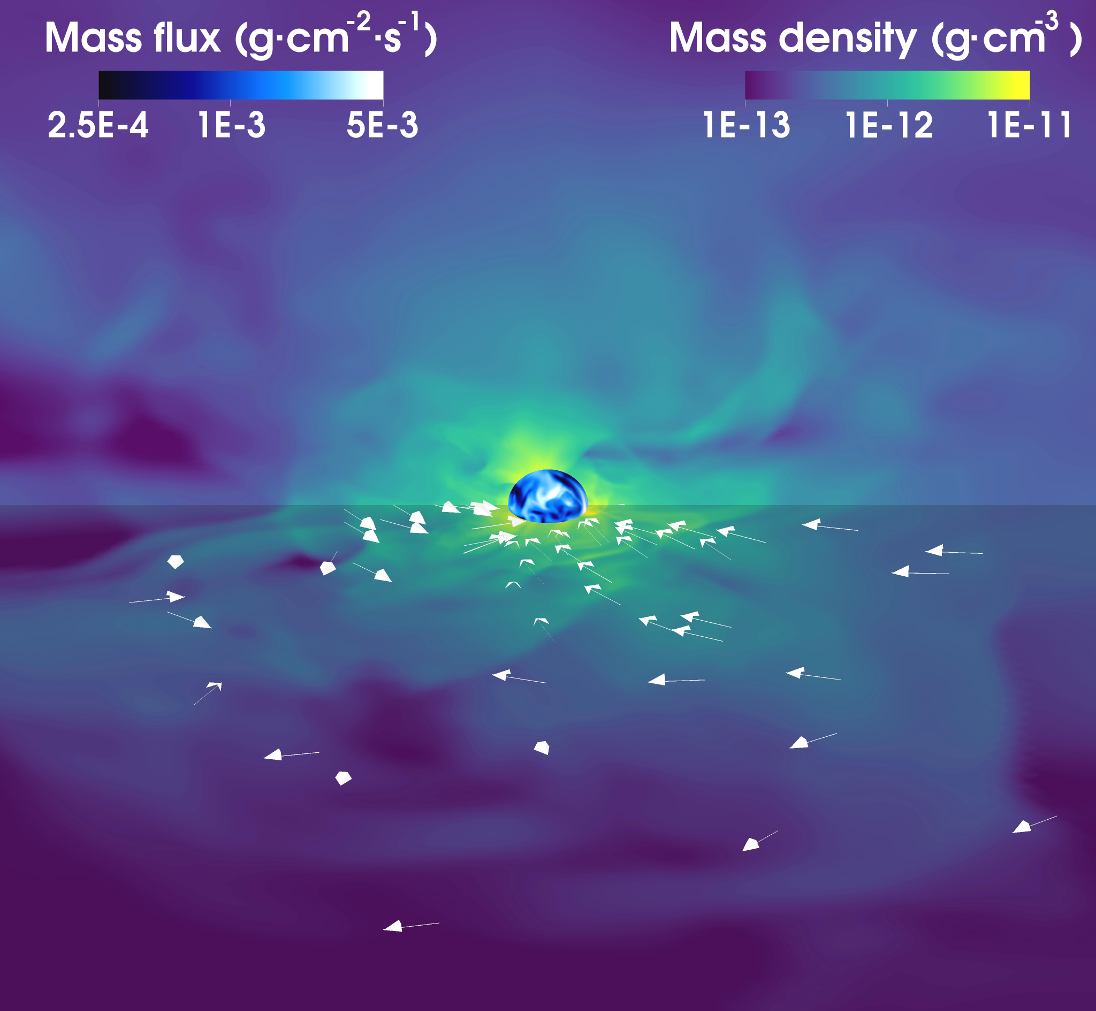}
     }
     \vspace*{-0.3cm}
     \caption{Side view on slices of a mass density colormap. The supersonic flow comes from the left and the polar axis is vertical. In the left panel, the solid white line is the Mach-1 contour. The black frame stands for the dimensions of the right panel. In the right panel, we zoom and show the velocity field in the equatorial plane. The surface of the inner sphere, of radius $R_\mathrm{acc}/100$, is colored with the absolute local radial mass flux.}
     \label{fig:side-view}
\end{figure*}

\subsection{Preliminary relaxation}
\label{sec:relax}

We first work with a planar and steady inflow to reach a numerically relaxed state and guarantee that the time variability measured in the simulations is fully induced by the inhomogeneities in the flow. The main computational cost of the simulations comes from (i) the large number of cells due to the full three-dimensional geometry and (ii) the large number of time iterations required due to the large contrast between the inner boundary and the outer boundary, a few hundred times larger. To alleviate this difficulty and reach a numerically relaxed state at an affordable cost, we proceed in the following way. First, we work on a two-dimensional radially stretched $\left(r,\theta\right)$ spherical grid similar to the one used in \cite{ElMellah2015} with an inner boundary of radius $r_\mathrm{in}=1/30$ (here and henceforth, length are in units of $R_\mathrm{acc}$). It is a larger inner boundary than the final one of $r_\mathrm{in}=1/100$, to reduce the number of time steps required to relax, but small enough to not significantly alter the detached bow shock formation. Then, the obtained axially-symmetric relaxed state is resampled on a three-dimensional spherical grid, rotated such that the polar axis is now orthogonal to the direction of the inflowing wind. We let this configuration relax in 3D and then, reduce the inner boundary radius down to its final value, $r_\mathrm{in}=1/100$. After a last stage of relaxation, we obtain a steady flow where the inner mass accretion rate varies at a level of less than 5\%. We can now set the inhomogeneous and time-varying upstream outer boundary conditions described in Section\,\ref{sec:outBC} and let the simulations reach a statistically steady state.

\begin{figure*}
\centering
\includegraphics[width=2\columnwidth]{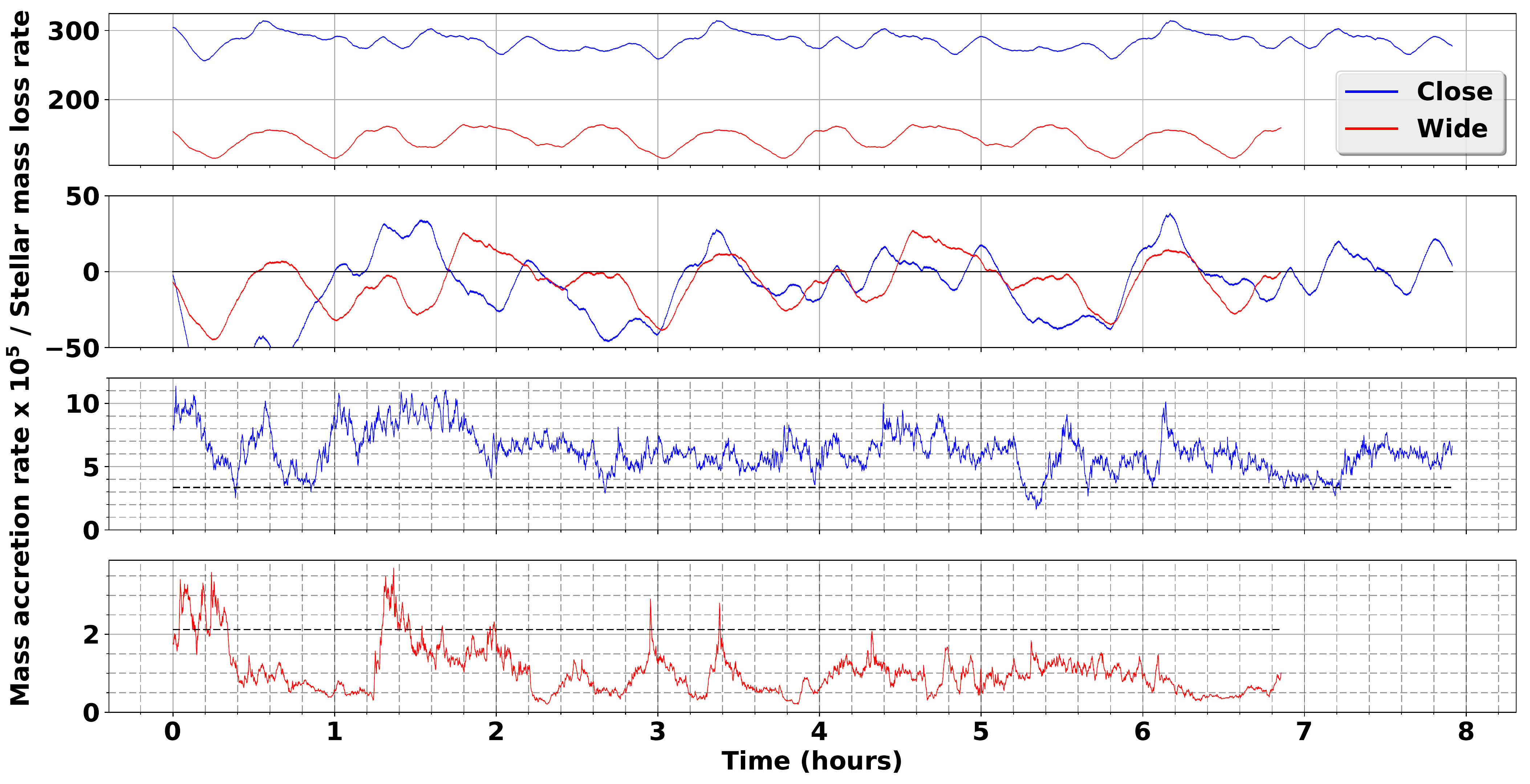}
\caption{Inflowing mass rate from the outer upstream hemisphere (top panel) and through the whole simulation space outer boundary (second panel), as a function of time, a few dynamical time scales after the first clumps reached the accretor. $\dot{M}$ at the inner boundary of the simulation space for the close ($a=1.6$R$_{*}$, third panel) and wide ($a=2$R$_{*}$, bottom panel) configurations. As a guideline, we plotted the BHL mass accretion rate for a smooth wind with the parameters from Table\,\ref{tab:param} (dashed black line). The mass flow rates have been scaled to the stellar mass loss rate divided by $10^5$.}
\label{fig:mdot_time}
\end{figure*}

\begin{figure}
\centering
\includegraphics[width=1\columnwidth]{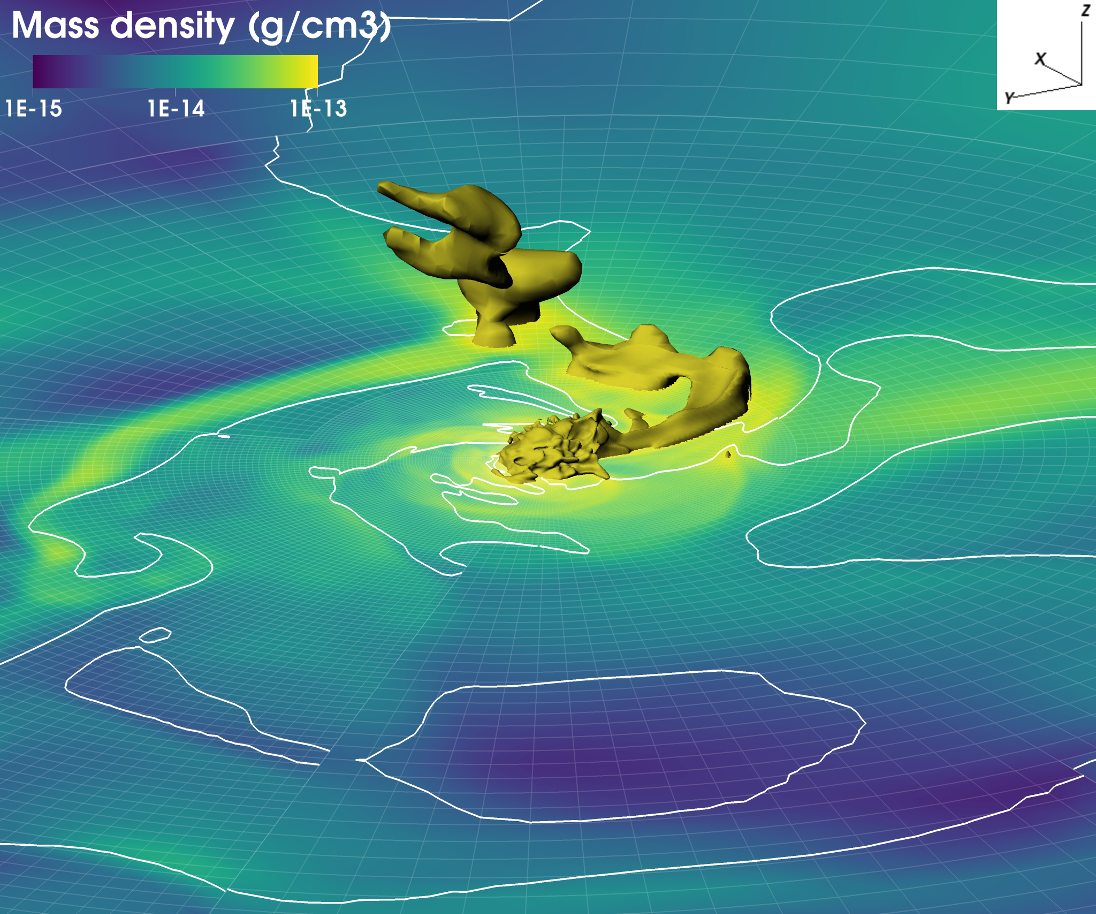}
\caption{Colormap of the density in the equatorial plane, from $10^{-15}$g$\cdot$cm$^{-3}$ to $10^{-13}$g$\cdot$cm$^{-3}$, with a three-dimensional isosurface of the density at $10^{-13}$g$\cdot$cm$^{-3}$. The solid white line is the Mach-1 surface in the equatorial plane which extends up to approximately 3 accretion radii (to the bottom left corner). The accretor lies in the center and in the upper right corner are drawn the axis : the wind comes from the direction $\hat{y}$ and the polar axis is $\hat{z}$. An overshooting clump is visible. Wide-orbit configuration ($a=2R_{*}$).}
\label{fig:clump_intruder}
\end{figure}

\section{Results \& discussion}
\label{sec:res}

\subsection{Structure of the flow}

The structure of the flow follows qualitatively what has been computed with a planar homogeneous inflow \citep{ElMellah2015}. As the supersonic flow approaches the accretor, it is deflected towards its wake and forms a bow shock with a distance between the shock front and the accretor of approximately one accretion radius (see left panel in Figure\,\ref{fig:side-view}). The inhomogeneities perturb the wake (see Figure\,\ref{fig:tail}). Also, the front shock moves back and forth but never reaches the inner boundary : the shock always remains detached.

Contrary to what has been reported from two-dimensional simulations, we do not see any significant flip-flop instability in the wake \ie a wobbling motion of the instantaneous direction of the trailing flow. Historically, this instability has first been witnessed by \cite{Matsuda1987} in two-dimensional numerical simulations and later confirmed and explored in more detail for different parameters \citep[see \eg][]{Blondin2009}. However, three-dimensional simulations proved way more stable when precautions concerning the geometry of the grid and the size of the inner boundary were taken \citep{Blondin:2012vf}. Since we do not consider any orbital effect in the current paper, we cannot exclude the possibility that this would make a flip-flop instability in the orbital plane possible. Within one accretion radius from the accretor, we do witness transient disc-like structures (see Figure\,\ref{fig:clump_intruder}) but the alignment of their angular momentum with the polar axis suggests that they are essentially due to the numerical setup ; indeed, the planar BHL problem does not admit any preferential transverse direction but our present numerical setup introduces one, both because of the polar axis and because of the 3D reconstruction method (see Appendix\,\ref{app:3Drec}). 

Concerning the wake itself, we measure a larger systematic opening angle of the bow shock compared to the steady homogeneous inflow configuration at high Mach number (see Figure\,\ref{fig:tail}). We see overdensities in the wake flowing away from the accretor. Their density contrast is much lower than in the upstream wind. Those overdensities form when a clump with an impact parameter too large (or, equivalently, with too much angular momentum) to be accreted reaches the shock, and propagates along the hull of the wake. 


\subsection{Mass accretion rate variability}
\label{sec:mdot}

We compute the mass accretion rate at the inner border $\dot{M}$ (2 bottom panels in Figure\,\ref{fig:mdot_time}), the mass flow rate entering through the outer upstream hemisphere (top panel) and the total net mass rate through the whole simulation outer border (second panel). The last serves as a proxy to estimate the moment when the simulation has reached a statistically steady state : it happens once the wind has crossed the whole simulation space, after approximately one hour. Once the numerically relaxed state for a homogeneous inflow has been reached (see Section\,\ref{sec:relax}), the mass accretion rate is constant to a 5\% level or less.

Since clumps are typically 20 times smaller than the diameter of the simulation space, the inflowing mass rate through the outer upstream hemisphere remains fairly constant as a function of time (top panel in Figure\,\ref{fig:mdot_time}). At a larger distance from the stellar surface, the clumps are larger and since the two configurations we consider have approximately the same accretion radius, it means that the clumps are comparatively larger for an accretor farther away from the star (wide-configuration, red line in Figure\,\ref{fig:mdot_time}). Hence a larger relative variability of the mass inflow rate, up to 20\%. As a comparison, we compute the BHL mass accretion rates for a smooth wind :
\begin{equation}
\dot{M}_\mathrm{BHL}=0.8\pi R_\mathrm{acc}^2 \rho_{\infty} v_{\infty}
\end{equation}
where $\rho_{\infty}$ is the density at the orbital separation (for an isotropic wind) and the 80\% efficiency factor is measured during the preliminary relaxation steps (see Section\,\ref{sec:relax}) and in 2D simulations \citep{ElMellah2015}. We evaluate $\dot{M}_\mathrm{BHL}$ (dashed lines in bottom two panels in Figure\,\ref{fig:mdot_time}) using the parameters from Table\,\ref{tab:param} and they approximately match the order of magnitude of the time-averaged $\dot{M}$ measured. However, such smooth-wind BHL estimates fail to capture the time evolution of the mass rate through the inner edge of the simulation space, once the flow has been shocked. Let us now discuss this time-variability. 

First, we examine the behavior of the flow at the shock. Sometimes, under the joint action of several clumps or of a particularly massive and cold one, the shock recedes down to the vicinity of the inner boundary. In this case, a clump can happen to get close to the inner boundary without undergoing much deformation. In Figure\,\ref{fig:clump_intruder} for instance, although the front shock, delimited by the Mach-1 contour (solid white line) stands firmly along the inflowing axis $\hat{y}$, it got breached on its side, at a distance of approximately one accretion radius from the accretor, by a stretched but unshocked clump. However, this clump has a slightly too high impact parameter and the angular momentum it carries is so important that it is not fully accreted ; it overshoots the accretor. Furthermore, at the same time, the other side of the shock is fairly unperturbed, with little inflow, and cannot feed angular momentum of opposite sign to facilitate accretion. This is a key-feature of inhomogeneous BHL accretion : since the axisymmetry is broken, accretion is much less efficient, even within the accretion cylinder, because a laterally inflowing clump with no opposite axisymmetric clump to compensate for its angular momentum will not be accreted. It also explains why, when the first clumps reach the accretors, we saw a sharp drop in $\dot{M}$ in all numerical setups. Matter piles up at the shock and in the wake but accretion itself is triggered only when the inflowing angular momentum changes sign. The most important flares are triggered by the conjunction between a series of clumps of low impact parameters. Then, the front of the shock recedes, the overdensities penetrate deep and contrary to the case above of high impact parameter, they are directly accreted, leading to a flaring $\dot{M}$ by a factor of $\sim10$ within a few $10$ minutes. As a consequence, the time-variability is enhanced with respect to the one of the mass inflowing rate from the upstream hemisphere at 8$R_\mathrm{acc}$, but not as much as expected from the BHL argument where dissipation is idealized. 

An important qualitative conclusion to draw from the simulations is that an incoming large clump does not straightforwardly yield a flare in $\dot{M}$. Conversely, it means that much caution should be taken when we try to trace back the origin of a flare to a clump mass. Between the ballistic clumps and the X-ray emission regions lie many different regions which dilute the clumps, store matter and trigger their own instabilities and gating effects : time variability can be either enhanced or lowered depending on the physical conditions encountered by the flow as it gets accreted. There is no direct one-to-one relation between the time variability at the smaller scales and the one induced by the inhomogeneous wind. Matter can pile up for quite long in intermediate regions (\eg the shock or the corotation radius) before an instability which makes accretion possible is triggered. 


\subsection{Column density}
\label{sec:NH}

\begin{figure*}
\centering
\includegraphics[width=2\columnwidth]{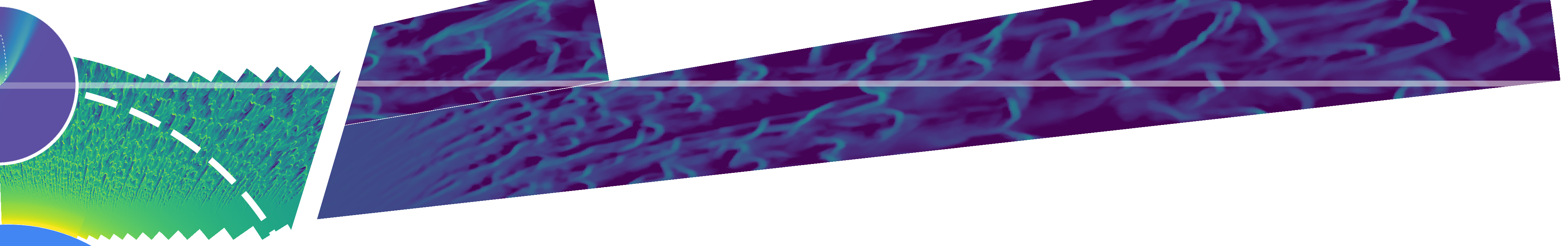}
\caption{Sketch (not on scale) of the computation of the time-varying $N_\mathrm{H}$, both inside the simulation space (represented in the upper left insert) and in the extended clumpy wind, at $\Phi_\mathrm{orb}=0.25$. For convenience, only the innermost ($<3$R$_{\odot}$) and outermost ($\lesssim$15R$_{\odot}$) regions have been represented, separated by a white gap. The LOS, going from the X-ray source up to a distance of $\sim15R_{*}$, is the semi-transparent white line (the observer lies to infinity towards the right). The NS orbit is the dashed white arch centered around the star (in the lower left corner). The central section has been masked. Similar scale as in Figure\,\ref{fig:snapshot} for the logarithmic density colormap in the outer region (arbitrary scale for the inner region in the upper left corner).}
\label{fig:sketch_NH}
\end{figure*}

The column density $N_\mathrm{H}$ is the integrated number density along the line-of-sight (LOS) :
\begin{equation}
N_\mathrm{H}=\int_{\text{LOS}}n \cdot dl
\end{equation}
with $n$ the number density of particles. Throughout this paper, we relate $n$ to the mass density in the simulations by assuming a mean molecular weight of 1. Once the opacity is known, $N_\mathrm{H}$ can be used to evaluate the absorption along the LOS at a given wavelength.

\subsubsection{Preliminary computation}
\label{sec:prelim}
We evaluate the time-varying $N_\mathrm{H}$ for an edge-on system using the following process. We put a constant X-ray point source at 1.8$R_{*}$ from the stellar center, embedded in the clumpy wind described above (see Figure\,\ref{fig:sketch_NH}). We work at orbital phases between $\Phi_\mathrm{orb}=0.2$ and $0.3$ (when the NS moves towards us). Then, we differentiate two regions :
\begin{itemize}
\item the simulation space, where we can compute on-the-go $N_\mathrm{H}$ in 256 directions in the equatorial plane, synchronously with the mass rates in Figure\,\ref{fig:mdot_time}.
\item the region centered on the stellar center and extending up to $\sim15R_{*}$, deprived of the simulation space (henceforth "the outer region").  
\end{itemize} 

To reconstruct the wind in the outer region from the stripe displayed in Figure\,\ref{fig:snapshot}, we convert the 2D pseudo-planar coordinates into proper polar ones. Then, we stick together, next to each other, stripes corresponding to different time snapshots, so as to laterally extend the signal. If we assume a self-similar extension of the clumps beyond $2R_{*}$, we can now extrapolate the results up to $15R_{*}$ by applying a homogeneous radial stretching factor to each stripe, computed such as the LOS of the observer (see semi-transparent straight line in Figure\,\ref{fig:sketch_NH}) intersects the stripe. To scale the mass density so as to guarantee a conserved mass loss rate at all radii, we fit the velocity profile computed in the LDI wind simulations (between $R_{*}$ and $2R_{*}$) with a $\beta$-law \citep{Castor1975} :
\begin{equation}
v\left( r \right)=v_{0}\left( 1-\frac{R_*}{r} \right)^{\beta} \quad \text{where} \quad \left\{
    \begin{array}{ll}
        v_0\sim 2,240 \text{km}\cdot\text{s}^{-1} \\
        \beta \sim 0.89
    \end{array}
\right.
\end{equation}
and assume that this profile is valid up to $15R_{*}$. $r$ is the distance to the stellar center, $v_{0}$ the terminal speed of the wind and $v_{0}$ and $\beta$ are the parameters we fit for. Since we work around $\Phi_\mathrm{orb}=0.25$, the LOS does not sample regions too distorted by this stretching. We computed $N_\mathrm{H}$ along the LOS in this outer region as the NS orbits from $\Phi_\mathrm{orb}=0.2$ to $\Phi_\mathrm{orb}=0.3$ and as the clumps move outwards. Beyond the expected systematic decrease in column density as the compact object moves towards us, we see a peak-to-peak spread of approximately $0.6\cdot 10^{22}$cm$^{-2}$ and a correlation time of at most 1 kilosecond. It is longer than the self-crossing time associated to a single clump, and results in a larger $N_\mathrm{H}$ enhancement, because several independent clumps along the LOS can happen to act jointly. From now on, we use this value of $0.6\cdot 10^{22}$cm$^{-2}$ as an estimation of the variability of $N_\mathrm{H}$ in the outer region at any $\Phi_\mathrm{orb}$. As long as the LOS does not intercept neither the trailing flow in the wake of the NS nor the deep layers of the wind close to the photosphere, this estimate should hold since the micro-structure of the wind is not significantly altered beyond a few accretion radii from the NS (see Section\,\ref{sec:ion}). 


\subsubsection{Column density within the simulation space}

Concerning the column densities computed within the simulation space, the equatorial plane of our mesh is representative of any plane which contains the main axis of the shock (\ie the direction of the inflow, $\hat{y}$ in Figure\,\ref{fig:clump_intruder}). Indeed, the present model is axisymmetric in the sense that, contrary to the approach of \cite{ElMellah2016a}, the source terms do not break the axisymmetry (\eg no Coriolis force and thus, no bending of the stellar wind). This procedure enables us to capture, with a very good time resolution, the behavior of $N_\mathrm{H}$ between a hundredth of accretion radius (\ie the outer rim of the NS magnetosphere) and 8 accretion radii (of the order of a tenth of the orbital separation) around the compact object. The aim of this procedure is twofold : (i) evaluating the respective impact on the column density variability of ballistic clumps on one hand, and the inner structures within the shocked region on the other hand, and (ii) checking whether phases of high accretion rates are associated to phases of high $N_\mathrm{H}$.

As discussed in Section\,\ref{sec:3D_Jon}, the clumps injected into our accretion simulation space have quite low sizes and mass. Although their extension is not uniquely defined and their shape is not spherical, the overdense regions tend to occupy a zone of a fifth of the accretion radius and to weigh a few $10^{17}$g. In terms of $N_\mathrm{H}$, when an unperturbed clump happens to pass along the LOS between the observer and the X-ray point source, it translates into an enhanced $N_\mathrm{H}$ of the order of $10^{21}$cm$^{-2}$ during approximately 100 seconds, in agreement with what we measure in our simulations. 


\begin{figure*}
\includegraphics[width=0.89\textwidth]{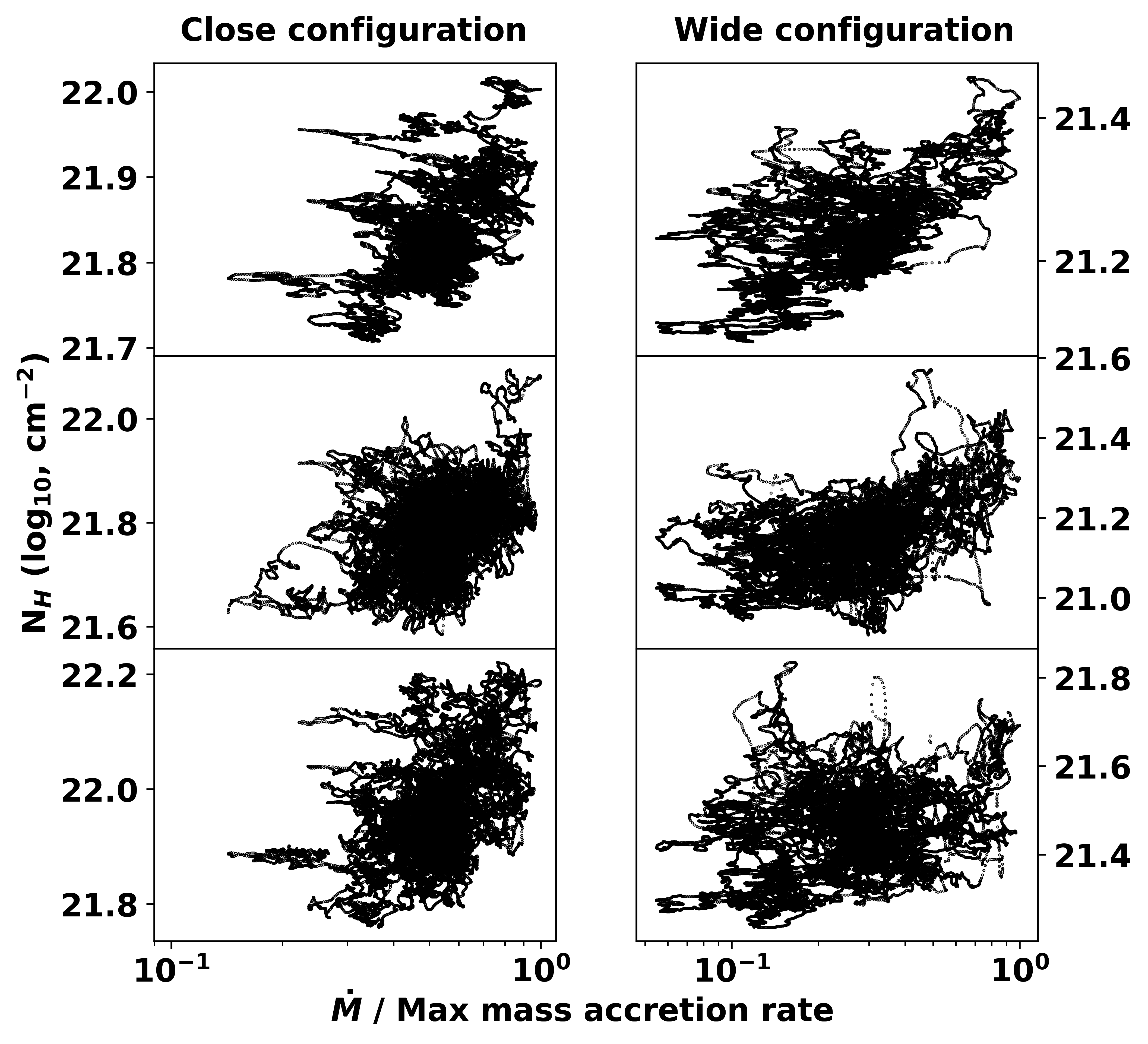}
\caption{Column density as a function of the full mass accretion rate at the inner border for the close (left column) and wide (right column) configurations. The top row is for column densities averaged in all directions. The middle row is the column density seen from the side (viewing angle in Figure\,\ref{fig:side-view} for instance). The bottom row is for column densities measured along the wake, in the back of the accretor.}
\label{fig:correl_mdot_NH}
\end{figure*}

Clumps can, on one hand, feed the accretor and on the other hand, raise the column density, whether they are accreted or not : a clump passing by the LOS but with an impact parameter much larger than the accretion radius will induce an enhanced absorption but not an X-ray flare. In Figure\,\ref{fig:correl_mdot_NH}, we plot, as a function of the inner mass accretion rate and for both configurations, the column density averaged in all directions (top panels), seen from the side (middle panels, at $\Phi_\mathrm{orb}=0.25$ or $0.75$ \ie same viewing angle as Figure\,\ref{fig:side-view}) or from the wake (bottom panels, $\Phi_\mathrm{orb}=0.5$ \ie inferior conjunction). A flare would be associated to an excursion to the right, sometimes associated to a rise in column density if the clumps responsible for this flare happened to pass by the LOS just before being accreted (see \eg the emerging loop in the upper right corner of the right middle panel). The upper panels display a slight positive correlation between the episodes of enhanced accretion and of larger quantities of mass integrated around the inner boundary, with Pearson correlation coefficients ranging from 0.3 to 0.7. However, for given viewing angles (\ie exposure times small compared to the orbital period), those results show that we should not observe a significant correlation between the column density and the X-ray luminosity, in agreement with the latest observations by \cite{Bozzo2017}.

\begin{figure}
\includegraphics[width=0.49\textwidth]{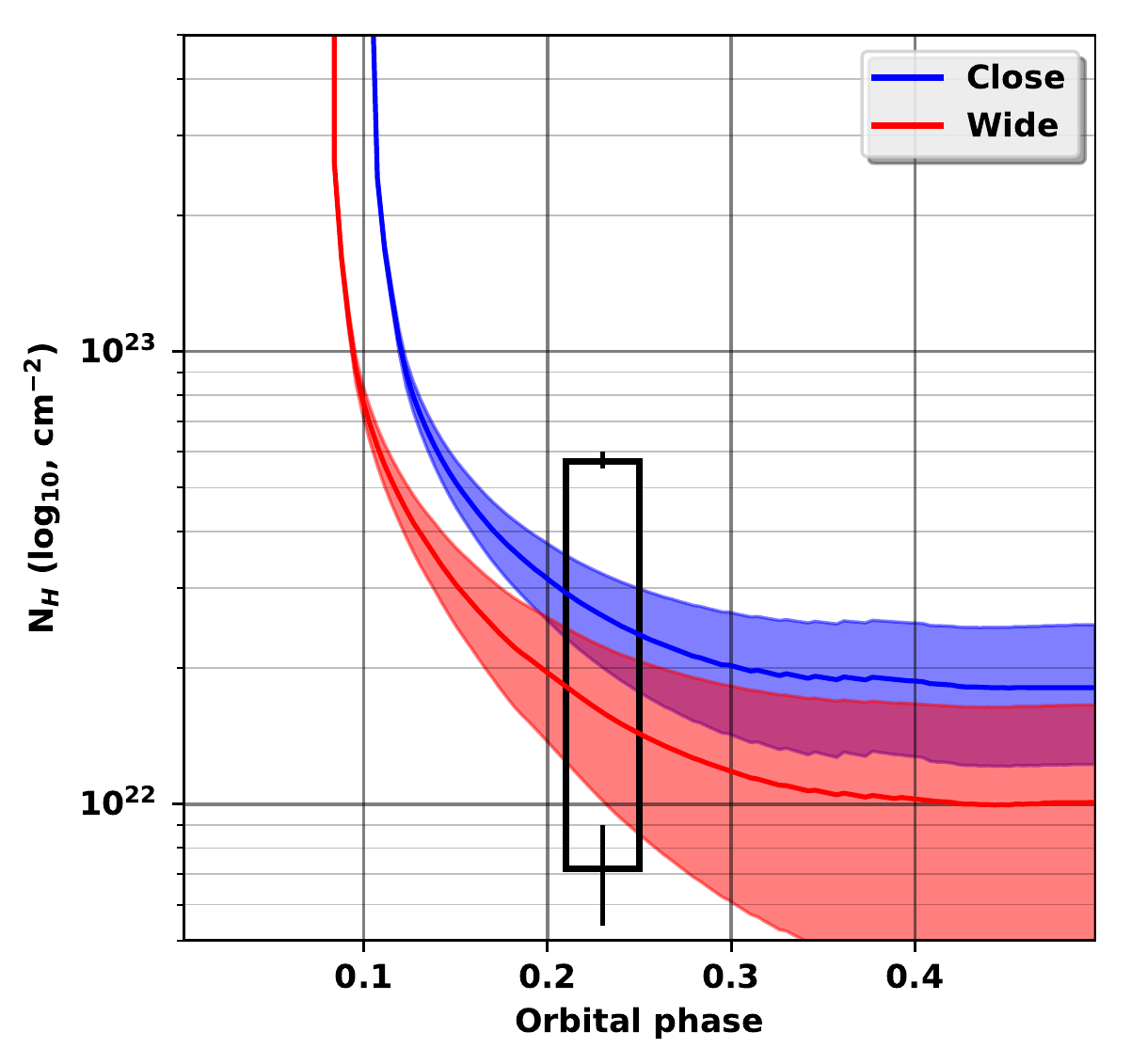}
\caption{Folded column density as a function of the orbital phase, for edge-on inclinations. The spread represents the 2-$\sigma$ time-variability of $N_\mathrm{H}$. The framed area shows the extent and the uncertainty on the variability observed by \citep{Grinberg2017}, at $\Phi_\mathrm{orb}=0.21-0.25$ (see text).}
\label{fig:NH}
\end{figure}

\subsubsection{Orbital column density of a smooth wind}

Our simulations last a few $10$ kiloseconds at most, which represent only a few percent of the total orbital period for a system like Vela X-1. If we suppose that the time variability of $N_\mathrm{H}$ measured within the simulation space is representative of its evolution at any $\Phi_\mathrm{orb}$, we can simply add the median value for each angle to the corresponding $N_\mathrm{H}$ as a function of phase computed from a smooth wind in the outer region. This smooth wind is the isotropic arithmetic and angular average in time of the clumpy wind model used throughout this paper. For the spreading, we use the quadratic sum of $0.6\cdot 10^{22}$cm$^{-2}$ obtained in Section\,\ref{sec:prelim} and the standard deviation computed in each  direction within the simulation space. It yields an estimate of the variability of $N_\mathrm{H}$ for each $\Phi_\mathrm{orb}$, along with a median value. 

The results are displayed in Figure\,\ref{fig:NH} for the close and wide configurations. Since we do not account for any non axisymmetric effect in the present paper, the results between $\Phi_\mathrm{orb}=0$ to $0.5$ are the same as the ones for $\Phi_\mathrm{orb}=0.5$ to $1$. On the left of the diagram is the eclipse region, at upper conjunction. Except in this region, the results for an edge-on inclination and a grazing inclination depart little from each other and thus, only the former is plotted. Due to the conservation of mass, the absolute values of the column densities scale approximately linearly with the stellar mass loss rate. Here, it is set to $1.3\cdot 10^{-6}$M$_{\odot}\cdot$yr$^{-1}$ in agreement with the value measured by \cite{Sako1999}, \cite{Watanabe2006} and \cite{Gimenez-Garcia2016}. Without surprise, we see that for larger distance of the accretor from the photosphere, the median $N_\mathrm{H}$ at any orbital phase is lower. Concerning the time variability relative to the median $N_\mathrm{H}$, represented by the colored area, it depends essentially on the size of the clumps with respect to the accretion radius : a larger orbital separation yields larger clumps and a smaller accretion radius (since the velocity of the wind is larger) while a smaller mass of the accretor gives a smaller accretion radius (see Equation\,\ref{eq:racc}). Both effects would increase the spreading around the median trend. As long as it remains small compared to the stellar mass, the mass of the accretor, $M$, has little influence on the average $N_\mathrm{H}$ profile. Within the statistically axisymmetric wake, $N_\mathrm{H}$ within and outside of the simulation space are comparable.

The black rectangle in Figure\,\ref{fig:NH} represents recent results on $N_\mathrm{H}$ in Vela X-1 at $\Phi_\mathrm{orb}=0.21-0.25$. During this interval, \cite{Grinberg2017} observed two periods of one and two hours where the hardness ratio increased dramatically. A spectral analysis of the continuum led them to attribute this change to a sudden rise of $N_\mathrm{H}$ from $0.72\cdot 10^{22}$cm$^{-2}$ to $5.76\cdot 10^{22}$cm$^{-2}$ (the upper and lower end of the black rectangle in Figure\,\ref{fig:NH}, along with the uncertainties). Both the close and wide configurations yield column densities lying within this range but they fail to explain this dramatic change in $N_\mathrm{H}$, which led \cite{Grinberg2017} to discard clumps happening to cross the LOS as possible explainations. Indeed, leaving aside the uncertainty on the median value (which can be shifted by tuning, for instance, the stellar mass loss rate), the present model is unable to reproduce the observed contrast in column density at these phases. It might be due to underlying structures within the magnetosphere that we could not study in detail with the present numerical setup. 


\subsection{Effective variability}
\label{sec:effective_mdot}


\begin{figure}
\captionsetup[subfigure]{labelformat=empty}
\centering
\subfloat[]{\includegraphics[width=0.5\textwidth]{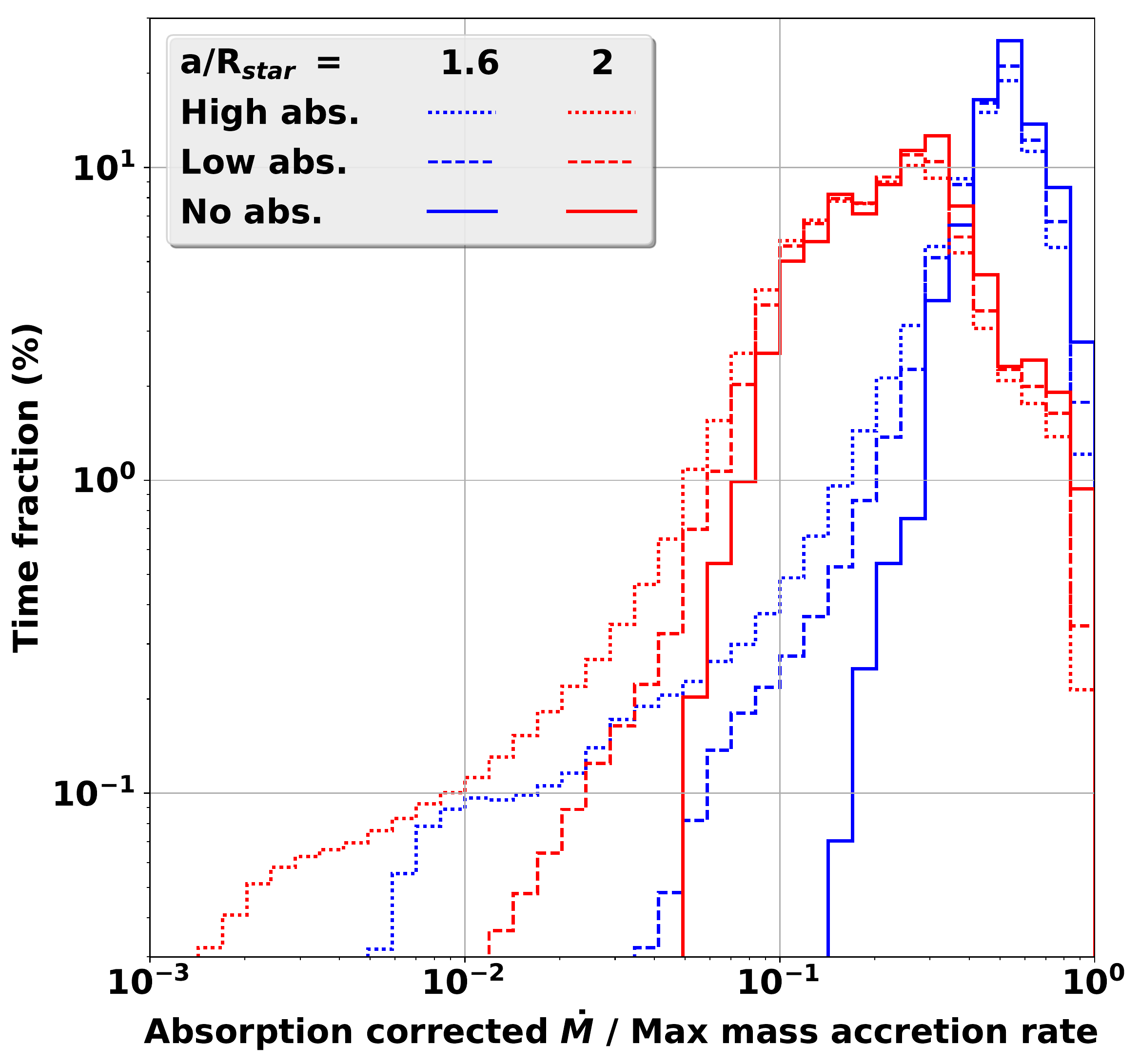}}
\vspace*{-0.5cm}
\subfloat[]{\includegraphics[width=0.5\textwidth]{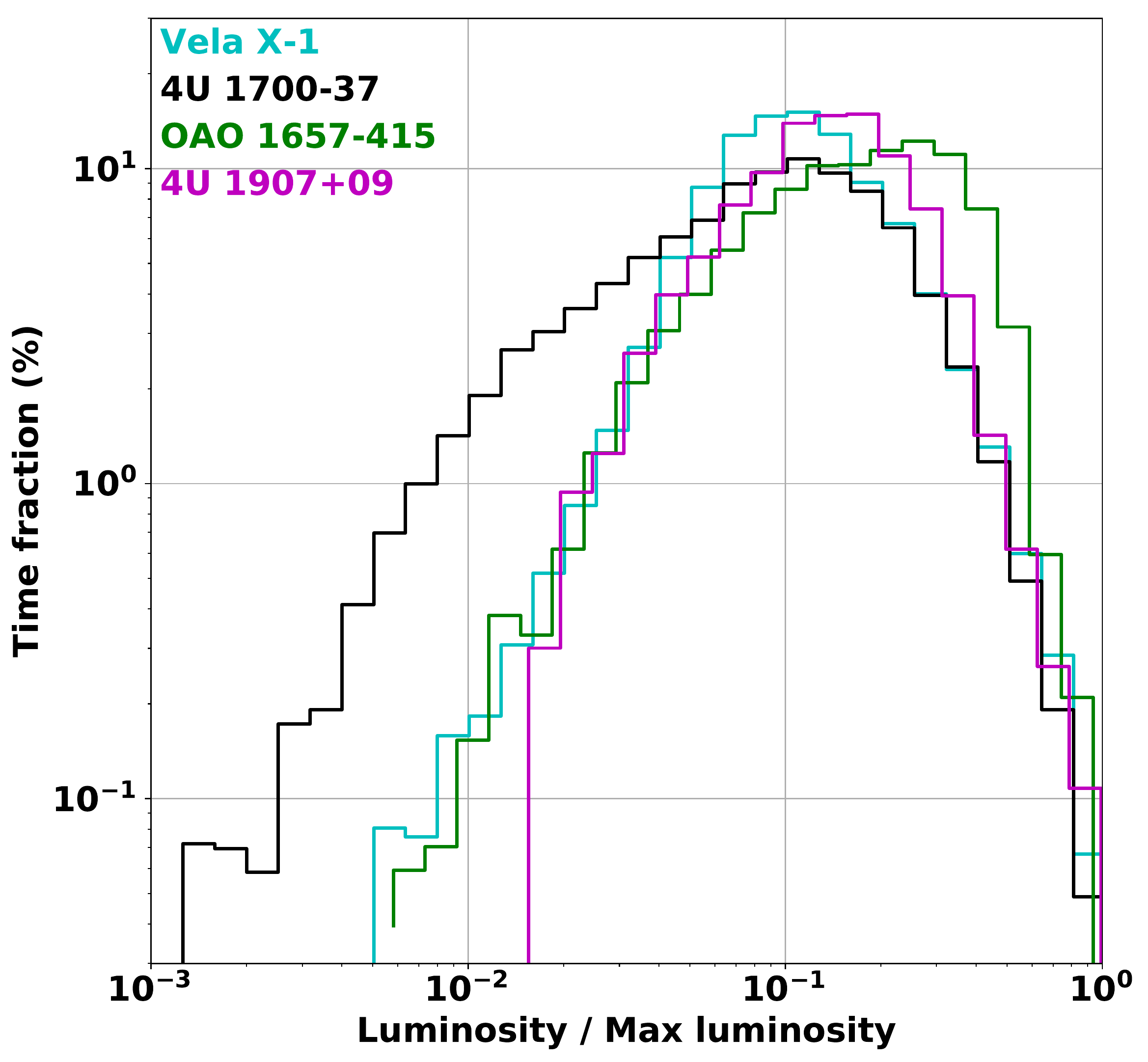}}\hfill
\vspace*{-0.6cm}
\caption{(upper panel) Simulated pseudo-luminosity histograms for both orbital separations seen edge-on. The low (resp. high) absorption case is for a photoionization cross section of $10^{-23}$cm$^2$ (resp. $2\cdot 10^{-23}$cm$^2$). (bottom panel) Luminosity diagrams for classic \sgx. The luminosity is compared to the maximum luminosity observed in each of those systems \citep[from][]{Walter15}.} \label{fig:diag}
\end{figure}

The upper panel of Figure\,\ref{fig:diag} shows, with solid lines, the histogram of $\dot{M}$ at the inner boundary compared to its maximal measured value. The time fraction represents the time spent by the system at this level of accretion, compared to the physical simulation duration ($\sim$8h). We discard the eclipses in this section. To guide the reader and discuss the results, we also show the observed luminosity diagrams for 4 classical \sgx in lower panel of Figure\,\ref{fig:diag} : Vela X-1, 4U 1700+37, OAO 1657-415 and 4U 1907+09. These results are extracted from Figure 8 of \cite{Walter15} where the eclipses and the off-states have been clipped off and the systems have been observed with Swift/BAT, in an energy range where little absorption is expected.

The first comment to make is that without absorption (solid lines in upper panel of Figure\,\ref{fig:diag}), we retrieve approximately log-normal distributions \citep{Furst2008,Furst2010,Walter15}, although cut at high activity. A serious discrepancy with observations concerns the restricted time-variability we see without absorption : for the close-configuration, $\dot{M}$ varies by a factor of approximately 10 while for the wide-configuration, this factor is $\sim20$. The time-variability of the hard X-rays emission in those systems is somewhat higher, between a few 10 and 100. It might be partly due to the limited duration of the simulations compared to the orbital period which might not have let the time to more extreme events (at the high and low ends of the activity level) to manifest. This can however not fully explain this discrepancy because the simulations of S17 did reach a statistically steady state, with a characteristic self-coherent time scale of the order of a few hundred of seconds at most. Consequently, we should already capture a representative variability of the inflow over the few hours of our simulations and should only miss marginally extreme events. The small scale clumps considered here are thus unlikely to fully explain the high time-variability observed in these systems, even on longer duration than 8 hours. Note however that the 2D LDI simulations used here are quite likely to somewhat underestimate the quantitative level of clumping, and moreover that the model has been computed for 
a generic OB-supergiant in the Galaxy and not been specifically tuned to the parameters of Vela X-1 (see discussion below). Other potential mechanisms for enhancing the time-variability include varying X-ray ionizing feedback at the orbital scale \citep{Manousakis2015c}, magnetic-centrifugal gating at the outer rim of the NS magnetosphere \citep{Illarionov1975,Bozzo2008} or quasi-spherical subsonic settling accretion \citep{Shakura2017}. Those mechanisms would possibly enable larger quantities of matter to be placed in "stand by" before being suddenly accreted once the physical trigger is matched. On the other hand, given their dimensions and their interaction with the bow shock surrounding the accretor, the clumps are not able to be stored for so long and flushed so suddenly. As a consequence, the wind inhomogeneities must be larger than those considered here in order to produce larger time-variability than a factor of 20 in $\dot{M}$ at approximately 1,000 NS radii from the accretor.

But is this relatively low contrast between high and low mass accretion regimes in our simulations due to a lack of high or low activity levels? The absolute values of $\dot{M}$ tend to favor the former option. Indeed, in the case of the close-configuration, the maximum accretion luminosity (with a maximum accretion efficiency of 1) obtained in Figure\,\ref{fig:mdot_time} amounts to $7.5\cdot 10^{36}$erg$\cdot$s$^{-1}$. For the wide-configuration, it hardly reaches $2.5\cdot 10^{36}$erg$\cdot$s$^{-1}$. Since the conversion of $\dot{M}$ into X-ray emission is generally considered to have an efficiency of a few 10\% \citep{Bozzo2008} and that the peak luminosity in a system like Vela X-1 is closer to a couple of $10^{37}$erg$\cdot$s$^{-1}$, we likely underestimate the maximum $\dot{M}$ by an order of magnitude. In the close-configuration, a 2 solar masses accretor and a 30\% slower wind \citep{Sander2017} could explain this discrepancy. Since the time-variability depends essentially on the size of the clumps comparatively to the accretion radius, clumps 4 times larger would be required to display similar levels of time-variability. In this respect, it is important to point out here that the LDI wind models considered in this paper deliberately stabilize the wind base at the stellar photosphere (see discussion in S17). While this provides a perfect, controlled environment for the present pilot-study of clumpy wind accretion, it also somewhat underestimates the level of clumping. Indeed, in the preliminary rotating 2D LDI-models by S17, the small-scale clumps are embedded in larger-scale structures emerging directly from the stellar surface. Future work will examine if considering these structures will bring up the the time-variability to the observed levels. Finally, more realistic lower wind speeds for systems like Vela X-1 \citep{Gimenez-Garcia2016} would increase the accretion radius but the altered size of the clumps it would be associated to remains to be studied to make any conclusive statement about the relative size of the clumps compared to the accretion radius and, by then, about the amplitude of the time-variability of the mass accretion rate.

In any case, the frequency of high activity levels is indeed underestimated in the present simulations. On the other hand, the low activity levels are captured accurately and match the ones of Vela X-1, OAO 1657-415 and 4U 1907+09, once we assume maximum $\dot{M}$ an order of magnitude higher (which would shift the histograms to the bottom left in the upper panel of Figure\,\ref{fig:diag}). Overall, this suggests that the time-variability observed at low luminosity in \sgx is essentially due to clumps generated in the line-driven wind of the donor star, whereas further studies are required to determine if this could be true also for the high-luminosity events. 

Even if we do not expect significant absorption in the photo-energy ranges observed in the bottom panel in Figure\,\ref{fig:diag}, we still evaluated the impact of absorption on these results, for illustrative purposes. In the soft X-rays ($\sim1$keV) though, the emission is strongly absorbed and the column densities previously measured could not be discarded. Absorption would naturally account for the asymmetries in the aforementioned log-normal distributions, with an enrichment of the low luminosity end. To compute a time-variability from the X-ray emission which accounts for absorption, we rely on the time and phase-dependent $N_\mathrm{H}$ within the simulation space and on the $\Phi_\mathrm{orb}$-dependent $N_\mathrm{H}$ out of the simulation space, computed from a smooth wind (see Section\,\ref{sec:NH}). Notice that due to the size of the inner border of our simulation space, we do not account for the column density between one and a few hundreds times the accretor radius. Also, the column densities we computed are for a stellar mass loss rate of $1.3\cdot 10^{-6}$M$_{\odot}\cdot$yr$^{-1}$ and the uncertainty reported in the literature around this value is at least a factor of 2 \citep{Sako1999,Watanabe2006,Gimenez-Garcia2016}. $N_\mathrm{H}$ would scale as the mass loss rate, provided the velocity profile is unaltered. So as to span the full half orbital period (from inferior to superior conjunction), typically a few days, we repeated by concatenating the $\dot{M}$ function of time displayed in Figure\,\ref{fig:mdot_time}, with random initial phases. We also work with an edge-on inclination. To visualize the impact of those uncertainties, we consider low and high absorption configurations with a photoionization cross-section of respectively $10^{-23}$cm$^2$ and $2\cdot 10^{-23}$cm$^2$ \citep{Wilms2000}. We also compare those results to a zero-absorption case, representative of the hard X-ray emission (above 10keV). Once corrected for absorption, we obtain effective mass accretion rates or "pseudo-luminosities" and plot their histogram in upper panel of Figure\,\ref{fig:diag} (dashed and dotted lines). We see that when absorption is accounted for, the low activity part of the diagram is indeed enriched, while the high activity region is essentially unaltered.


%


\section{Summary \& outlook}
\label{sec:conc}

We ran 3D simulations of the accreted flow, centered on the accretor and with a mesh suitable to resolve the inhomogeneities in the upstream wind and follow them over almost 3 orders of magnitude in space. As the flow enters the sphere of influence of the NS, it is beamed toward it and forms a detached bow shock. The successive clumps cross this shock and, provided they loose enough energy and angular momentum, get accreted. As expected from the BHL approach of clump accretion \citep{Ducci2009}, the accretion proceeds mostly through the accretion cylinder. However, the present hydrodynamical simulations show that the shock lowers the time-variability with respect to the BHL estimation. Also, it introduces a time lag and a phase mixing since the shocked material associated to a clump is not straightforwardly accreted : it might be stored in a transient disc-like structure before accretion of matter with opposite net angular momentum triggers effective accretion. 

Comparing to the X-ray luminosity diagrams at high energy, we showed that the wind micro-structure computed by the clumpy wind models considered in this paper is not sufficient, per se, to retrieve the time-variability levels observed in classical \sgx such as Vela X-1, 4U 1700-37, OAO 1657-415 or 4U 1907+09 \citep{Walter15}. The behavior at low luminosity matches the observations but the largest luminosity events require the possibility to quickly tap larger amounts of matter the clumps we considered here are not able to provide, even accounting for the intermediate shocked region where the flow can pile up. Other storage stages can appear once the NS magnetosphere and radiative cooling is accounted for \citep{Illarionov1975,Bozzo2008,Shakura2017} or at the orbital scale \citep[\eg the X-ray ionizing feedback in][]{Manousakis2015c}. However, as discussed above, while the LDI simulations used here provide a perfect test-bed for this first study, such simulations do indeed yield larger clumps at a given orbital separation for a larger star and/or when clumping is enabled earlier on, near the stellar surface \citep[where we know that clumps are already present,][]{Cohen2011,Sundqvist2013,Torrejon2015}. Building on this first, generic study of 3D wind accretion from a line-driven clumpy wind outflow, future work will thus use LDI simulations more specifically tuned to Vela X-1 to examine in more detail how well the clumpy accretion can explain also the recent observations by \cite{Grinberg2017}.


%
%

\section*{Acknowledgments}

The authors are indebted to the anonymous referee who brought up several insightful questions and helped to improve this paper. IEM has received funding from the Research Foundation Flanders (FWO) and the European Union's Horizon 2020 research and innovation programme under the Marie Sk\l odowska-Curie grant agreement No 665501. IEM and JOS are grateful for the hospitality of the International Space Science Institute (ISSI), Bern, Switzerland which sponsored a team meeting initiating a tighter collaboration between massive stars wind and X-ray binaries communities. IEM also thanks Peter Kretschmar, Victoria Grinberg and Felix F\"urst for the fruitful discussions and the relevant comments they made on the present work. This research was supported by FWO and by KU Leuven Project No. GOA/2015-014 and by the Interuniversity Attraction Poles Programme by the Belgian Science Policy Office (IAP P7/08 CHARM). The simulations were conducted on the Tier-1 VSC (Flemish Supercomputer Center funded by Hercules foundation and Flemish government).





\bibliographystyle{agsm}
\begin{tiny}
\bibliography{clumpy_wind}
\end{tiny}




\appendix


\begin{figure}
\centering
\includegraphics[width=1\columnwidth]{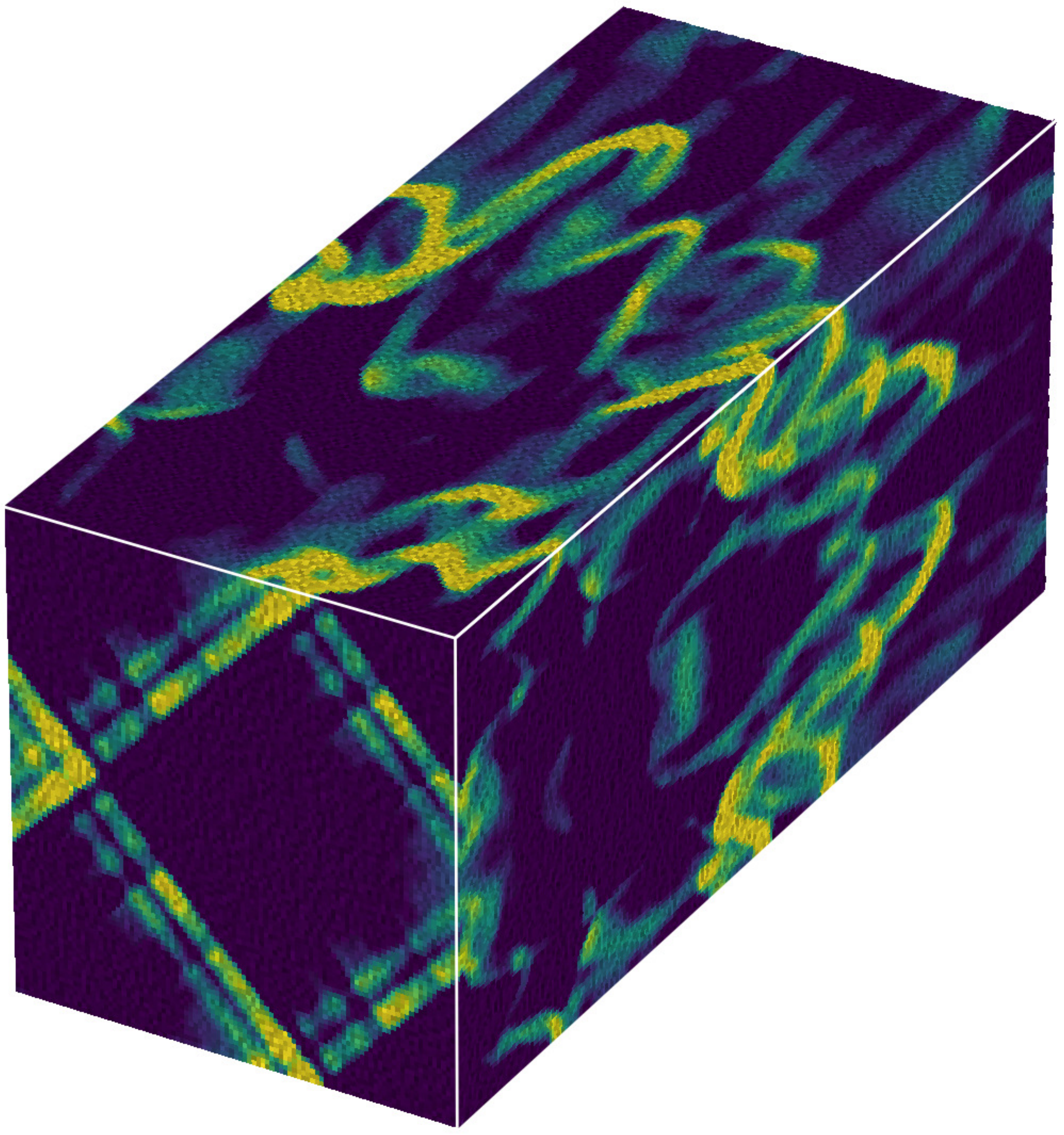}
\caption{Three-dimensional reconstructed mass density map at the orbital separation (aspect ratio respected). The front slice is transverse to the radial direction and the upper slice is the original signal from S17. The top/bottom symmetry can be seen on the right and front slices.}
\label{fig:3Dslice}
\end{figure} 

\section{3D reconstruction}
\label{app:3Drec}
To fully capture the dynamics of the flow around the compact object, three dimensional simulations of the accretion process centered on the accretor and spanning several orders of magnitude in space are required. Time varying outer boundary conditions are naturally provided by the 2D clumpy wind simulations presented in the main text, provided we develop a reconstruction method to expand these results along the second transverse dimension. In order to minimize the alteration of the results from S17, we use the following method. Let us write $s_{i,j}$ the information in each pixel of the 2D stripe at a given time (with $i$ for the radial direction and $j$ the transverse direction). Since the two transverse directions are equivalent, we compute a 3D signal $S_{i,j,k}$ by writing :
\begin{equation}
S_{i,j,k}=\overline{s_{i,j}\cdot s_{i,k}}
\end{equation}
where the over line indicates that we combine $s_{i,j}$ and $s_{i,k}$ either with a geometrical mean (for mass density and pressure) or with an arithmetic mean (for the radial velocity component). Concerning the transverse components of the velocity field, they are not retained in the present paper where we discard the shearing of the flow and the orbital effects. It yields the initial signal along the transverse diagonal ($\overline{s_{i,j}\cdot s_{i,j}}=s_{i,j}$) and a symmetric matrix for any index $i$ ($\overline{s_{i,j}\cdot s_{i,k}}=\overline{s_{i,k}\cdot s_{i,j}}$). We then rotate $S_{i,j,k}$ by 45 degrees around the radial axis to retrieve the initial signal in one of the two transverse direction and to make use of the aforementioned matrix symmetry and work only in the upper half of the simulation space. Finally, since the averaging procedure lowers the contrasts in the other transverse direction, we proceed to a histogram correction to retrieve approximately the same histograms for the mass density, the pressure and the speed in both transverse directions. 

The benefits of this approach are to conserve the transverse extension of the clumps, their longitudinal correlation length, their overall shape and to retrieve exactly the initial two dimensional signal along one transverse direction. On the contrary, the main drawback which must be acknowledged for is the spurious leakage along the two privileged orthogonal transverse directions associated to the matrices above : cross-shaped patterns appear in the transverse slices of the three dimensional reconstructed signals. Overall, this technique tends to produce spherical clumps \citep[see][]{Sundqvist2012}.

%

\subsection{Transverse prolongation}
\label{sec:trans_prol}
Since the width of the stripes ($R_*/10$) in the clumpy wind simulations turns out to be smaller than the required simulation space centered on the compact object (of radius $8R_\mathrm{acc}$, see Section\,\ref{sec:Racc}), we need to enlarge their transverse extension. Only a reduced fraction of the flow, the one with the smallest impact parameter ($\lesssim R_\mathrm{acc}$) with respect to the accretor, is expected to actually be accreted at some point \citep{ElMellah2015}. Beyond a few accretion radii, the flow must still be described within the simulation space but its precise micro-structure will not play a major role in the time variability of the inner mass accretion rate. It it thus legitimate to simply repeat the central three dimensional data in both transverse directions. Thanks to the periodic transverse boundary conditions used in the original wind simulations, this does not introduce any discontinuity.

\section{Numerical setup}
\label{app:num_divers}

\subsection{Multi-scale mesh}
\label{sec:mesh}

\begin{figure}
\centering
\includegraphics[width=1\columnwidth]{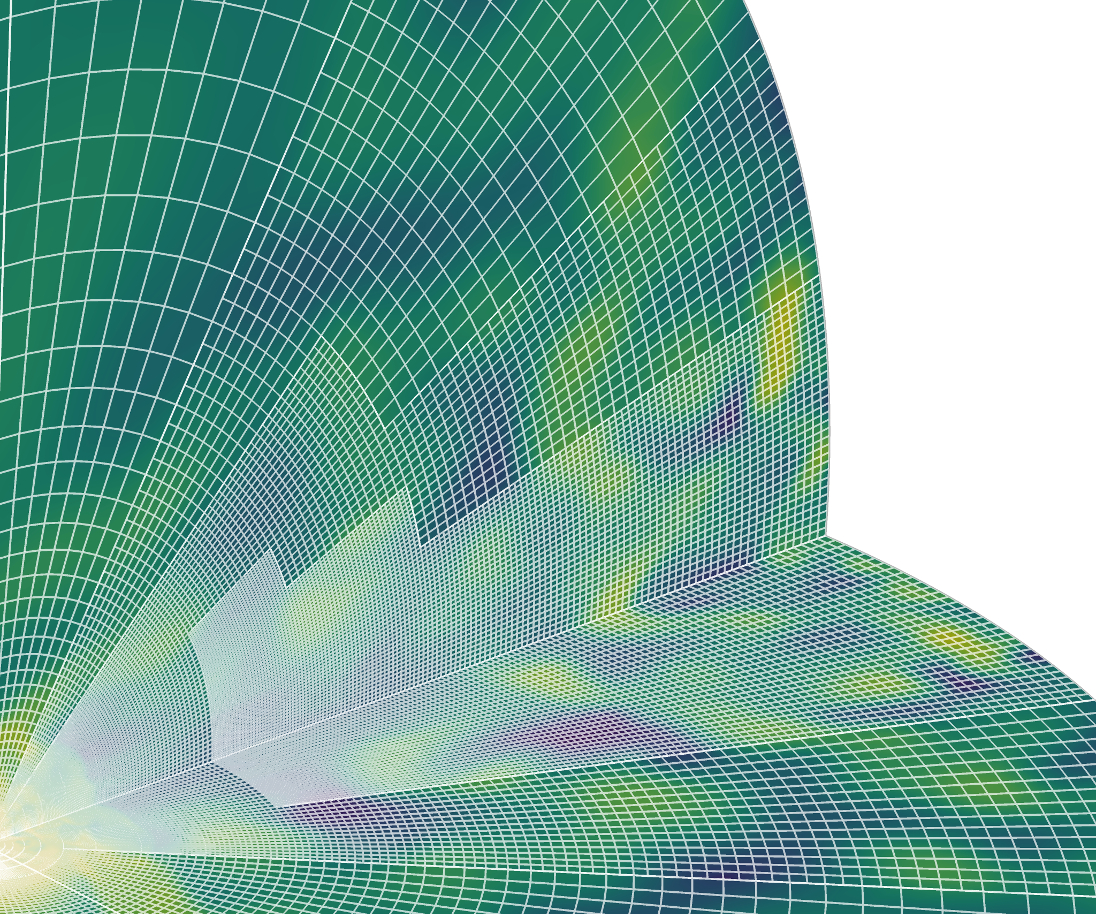}
\caption{Two slices of the upper upstream hemisphere of the radially stretched mesh we use. We overlaid a logarithmic density map to show the typical size of the inhomogeneities to resolve. As can be seen here, we prevented AMR around the pole and favored it in the accretion cylinder. The accretor lies in the bottom left corner, $8R_\mathrm{acc}$ away from the outer border.}
\label{fig:mesh}
\end{figure}

The mesh introduced here to afford numerical simulations of clumpy wind being accreted is partly displayed in Figure\,\ref{fig:mesh} and described in more detail in \cite{Xia2017}. The core physical requirements are (i) to be able to follow the flow from $8R_\mathrm{acc}$ down to the shock ($\sim R_\mathrm{acc}$) and below to resolve the sonic surface (a few hundredths of $R_\mathrm{acc}$) and (ii) to be able to resolve the wind micro-structure at any time, in particular in the regions susceptible to be accreted.

\subsubsection{Radially stretched}

To uniformly probe the flow all along its journey to the accretor, it is not possible to work with a constant radial step. We designed a radially stretched mesh \citep{ElMellah2015} to make it affordable and to keep the same cell aspect ratio from the inner boundary of the grid up to the outer boundary. Doing so, we could span almost 3 orders of magnitude in space. It enabled us to properly compute the shock formation (\ref{sec:Racc}) and to study the accretion process within the shocked region, down to an inner boundary radius of a hundredth of the accretion radius. Larger inner boundaries lead to a significant alteration of the shock which becomes attached to the inner boundary. In physical units, the spatial extension of this inner boundary is approximately 1,000 times the size of the compact object ; for a neutron star, it corresponds to the region where the magnetic field role can no longer be neglected and where the hydrodynamic framework we rely on in this paper would break down.

\subsubsection{The pole}

Due to the shrinking of the azimuthal spatial step towards the poles and to respect the Courant-Lewy-Friedrich condition, three dimensional spherical simulations suffer strong limitations on the time step at the pole. In \cite{Blondin:2012vf}, the authors intertwined two spherical meshes rotated by 90 degrees around one of the two equatorial axis, so as to avoid the polar regions of each mesh (the Yin-Yang approach). Here, we simply degraded the resolution around the pole by preventing any AMR to take place within a certain angular region around the pole. Doing so, the grid remains at its first resolution level, typically $64\times 48\times 64$, and refining the mesh in the equatorial regions does not entail smaller time steps up to an AMR level of 6 : the only additional cost when refining is the higher number of cells, which can easily be handled given the excellent weak scaling of \texttt{MPI-AMRVAC} \citep{Porth:2014wv}. The price to pay for such an approach is a lower resolution at the poles and a risk that artificial smoothing of the flow along the pole favors spurious disc formation in the equatorial plane of the mesh. We work with 4 different levels of AMR such that the effective resolution is of $512\times 384\times 512$.
  
\subsubsection{Selective AMR}

The clumps enter the spherically stretched mesh from the upstream hemisphere. Because those clumps are off-centered small scale structures, we need to enable AMR (up to 4 levels) in this hemisphere to resolve them, in particular at the outer edge of the mesh where the radially stretched cells have the largest absolute size on the first AMR level. On the contrary, because we are only interested in the fraction of the flow susceptible to be accreted by the compact object, we inhibit AMR refinement in the downstream hemisphere, except in the immediate vicinity of the accretor (below the stagnation point in the wake). We also prevent excessive refinement in the vicinity of the accretor (no refinement beyond the third level) since the stretching already provides more refined cells. Finally, in the upstream hemisphere, we favor AMR refinement in the accretion cylinder, around the axis of zero impact parameter. All those precautions enable us to follow the flow as it is accreted onto the compact object while still resolving off-centered inhomogeneities.


\subsection{Code and solving scheme}

We solve the adiabatic equations of hydrodynamics under their conservative form using the finite volume code \texttt{MPI-AMRVAC} \citep{Porth:2014wv}. The energy equation contains no cooling nor heating term, and the gravitational field of the compact object is the only source term we consider. We use a shock-capturing HLL solver \citep{Harten1983}, and a Koren slope limiter \citep{Koren1993}. The Courant number is typically set to 0.5. We also wrote a new user-defined type of boundary condition to work with time-varying boundary conditions stored in data files. A test-case of time-varying boundary conditions is now part of the main branch of the code.


\bsp	
\label{lastpage}
\end{document}